\documentstyle[11pt,epsf]{article}
\newcommand{\AmS}{{\protect\the\textfont2
  A\kern-.1667em\lower.5ex\hbox{M}\kern-.125emS}}
\newcommand{\av}[1]{\mbox{$\langle #1 \rangle$}}
\makeatletter
\@addtoreset{equation}{section}
\makeatother
\renewcommand{\theequation}{\thesection.\arabic{equation}}
\begin{document}
%
\hbox{}
\mbox{}\hspace{1.5cm}January 1993 \hspace{8.0cm} HLRZ-93-7\\
\begin{center}
\vspace*{2.5cm}
{\LARGE The Confined-Deconfined Interface Tension in \vspace{0.2cm}\\
       Quenched QCD using the Histogram Method}\\
\vspace*{0.7cm}
{\large B.~Grossmann, M.~L.~Laursen\\ }
{\small
        HLRZ,c/o Kfa Juelich, P.O. Box 1913, D-5170 Juelich,
                    Germany}\\
\vspace*{3.0cm}
\end{center}
\begin{abstract}
We present results for the confinement-deconfinement interface tension
$\sigma_{cd}$ of quenched QCD. They were obtained by applying Binder's
histogram method to lattices of size $L^2\times L_z\times L_t$ for $L_t=2$ and
$L=8,10,12\mbox{ and }14$ and various $L_z\in [L,\, 4\, L]$.
The use of a multicanonical algorithm and rectangular geometries have turned
out to be crucial for the numerical studies. We also give an estimate for
$\sigma_{cd}$ at $L_t=4$ using published data.
\end{abstract}
%
\section{Introduction}
\label{introduction}

At high temperatures  the structure
of strongly interacting particles is supposed to be quite different from
its low temperature form. The familiar hadron spectrum will be dissolved and
quarks and gluons will become the fundamental degrees of freedom
(``quark gluon plasma''). These two phases are
probably separated by a first order phase transition at a critical temperature
$T_c\approx 100-200\,\mbox{ MeV }$
(see \cite{gav89:histnp,fuk89:histnp,alv90:histnp,iwa92:histnp}).
The existence of this transition might have interesting consequences for
the nucleosynthesis in the early universe. One possible scenario which could
happen when the temperature of the universe is cooled down below $T_c$ is the
following~\cite{kaj86:basf,ful88:histnp,lan93:histnp}: First the universe
got  slightly supercooled because of the extra
interfacial free energy which is required for the generation of regions of
hadronic phase within the quark gluon plasma. When this so called
''nucleation'' occurred, the universe was reheated to $T_c$ because of the gain
in latent heat. Then the generated hadronic bubbles growed keeping the
temperature  constant  until they finally met. The situation was now
reversed: There
were bubbles of quark gluon plasma within the hadronic phase. The
temperature could not be maintained any more by the gain of latent heat und
was thus decreased so that also the remaining bubbles  hadronized. Since the
baryonic density in the quark gluon plasma may be much higher than in the
hadronic phase~\cite{alc89:histnp}, an inhomogeneous baryon number
distribution resulted from this condensation process.
If the scale of these
inhomogeneities  is between the diffusion length that protons moved
until nucleosynthesis set in  and the one of the neutrons, the baryon
number inhomogeniety transformed into an inhomogeneity  of the neutron
to proton ratio. This obviously has consequences for the element
synthesis~\cite{lan93:histnp,alc89:histnp}.

In the early stage of this development an important new length scale shows up:
The average distance $R_i$ between the hadronic bubbles after the universe had
been reheated to $T_c$. Assuming that this point was reached when the shock
waves emitted from the hadronic bubbles met, this distance has been calculated
to be~\cite{kaj86:basf,ful88:histnp,ban92}
\begin{equation}
R_i = 7\cdot 10^5 \, \frac{(\sigma/T_c^2)^{3/2}}
                          {(L/T_c^4)\,T_c^2[\mbox{MeV}^2]}
\label{eq:ri}
\end{equation}
where L is the latent heat of the transition and
\begin{equation}
\sigma_{cd}=\frac{F}{A\,T_c}
\end{equation}
is the reduced interface tension of an interface  between the
hadronic (``confined'') and the quark gluon plasma (``deconfined'') phase.
 $A$ is the  area of this interface, and $F$ is its free energy.
Thus the knowledge of the values of $\sigma_{cd}$ and $L$ would give the
scale for the inhomogeneities in the baryon number which has been generated
by the deconfinement phase transition. In order to obtain inhomogeneities which
are not washed out by the diffusion of the protons, $R_i$ should be at least of
the order of 0.5m~\cite{ban92}.

The latent heat has been measured in e.~g.~\cite{iwa92:histnp}. The interface
tension has been  determined numerically at the critical
temperature $T_c = 1/L_t$ where $L_t$ is the lattice extent in the euclidean
time direction. Essentially two different types of methods have been used.
In the first type coexistence of the
confined and the deconfined phase was enforced by keeping different parts
of the lattice at different temperatures or by applying an external field. This
breaks translation invariance and pins the interface at a certain position.
The properties of a pinned interface will in general be different
from the ones of the free interface one is interested in. To
extract the interface tension one should first perform the infinite volume
limit
and then turn off the temperature gradient or the external field. In practice
this is difficult because in a numerical simulation the lattice size
is necessarily limited. This causes finite size effects which must be
understood before one can extrapolate results reliably to the infinite volume
limit. Using this type of methods for $L_t = 2$ the Boston
group obtained $\sigma_{cd} = 0.12(2) T_c^2$ \cite{hua90:hist} while the
Helsinki group quotes $\sigma_{cd} = 0.08(2) T_c^2$ \cite{kaj91:hist}.
 Closer to the continuum limit for $L_t = 4$ the Boston group quotes
$\sigma_{cd} = 0.027(4) T_c^2$ \cite{bro92:hist}. But the validity of this
 method  has been questioned lately by the authors of
refs.~\cite{jan92:histnp,jan92b:histnp}. Using the same method as in
ref.~\cite{bro92:hist} for the two-dimensional seven-states Potts model one
finds numerical values which are  in disagreement with an  analytic result
(see ref.~\cite{bor92c:histnp}).

The second type of calculations is done without any pinning of the interface.
One approach of this kind makes use of the finite volume splitting of the
spatial transfer matrix spectrum. For $L_t=2$ one obtains
$\sigma_{cd}=0.139(4)\, T_c^2$~\cite{gro92b:histesp}. In order
to apply this method, the extension of the lattice in one spatial direction
($L_z$, say) has to be much larger than the corresponding tunneling
correlation length $\xi_t$. Since the latter increases exponentially with
the area $A$ of the interface one is restricted to rather small interface
areas. In contrast to this for the modified version of Binder's histogram
method~\cite{bin81:hist} which we are using in this work one has
to use lattices of extension  $\xi_b\ll L_z\ll\xi_t$ where $\xi_b$ is the
bulk correlation length.

In this paper we apply Binder's histogram method to rectangular lattices
($L_z>L$). We demonstrate how this eliminates the interfacial interactions
which had complicated previous studies. We give a simple consistency criterion
for the absence of such interactions and show that our data are well described
by a capillary wave model. This reduces the number of fitting parameters to two
(one of which is the interface tension). We obtain
$\sigma_{cd}=0.103(7)\,T_c^2$ which
is consistent with the older data but slightly lower than the more accurate
determination by the transfer matrix method. We will argue that this might be
due to the rather small interfaces used for the transfer matrix method.

Our method is described in section~\ref{interfacial}. For the numerical work it
was essential to use a multicanonical algorithm for quenched
$QCD$~\cite{gro92:histas}. This is described in section~\ref{multica}.
In section~\ref{results} we give our numerical results for the tunneling
autocorrelation times, the critical coupling $\beta_c$ and the interfacial
free energy. Finally we give our conclusions in section~\ref{conclusion}.

\section{The Interfacial Free Energy}
\label{interfacial}
\begin{figure}[t]
\begin{minipage}{0.49\textwidth}
\vspace*{-1cm}
\vspace{5cm}
\caption{ {
Schematic probability distribution for the order parameter. The dotted
line indicates the multicanonical distribution.
}}
\protect\label{fig:pmc}
\end{minipage}
\hfill
\begin{minipage}{0.49\textwidth}
\centerline{
	\epsfysize=4cm
	\epsfbox{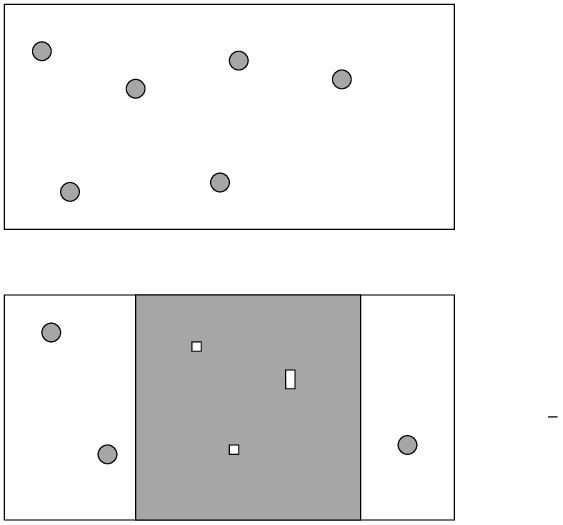}
	}
\caption{ {
Typical cuts through the lattice in the $y-z-$plane at values $\rho^{(1)}$
(first picture) and $\rho^{(min)}$ (second picture) of the order parameter. The
two phases are represented by white resp. shaded areas.
}}
\protect\label{fig:config}
\end{minipage}
\end{figure}

We consider SU(3) pure gauge theory with the Wilson action $S$ on rectangular
lattices of size $L^2\times L_z\times L_t$ and  $L_z$ varying between
$L$ and $4L$ at the critical coupling $\beta_c$ for $L_t=2$.
We use periodic boundary conditions in the time direction and $C-$periodic
boundary conditions \cite{kro91:hist} in the spatial directions, i.e.
\begin{eqnarray}
U_\mu(\vec{x} + L_i \vec{e}_i,t) &=& U_\mu^*(\vec{x},t),\mbox{ for }i=x,y,z\\
U_\mu(\vec{x} ,t + L_t) & = & U_\mu(\vec{x},t).
\end{eqnarray}
The value of the Polyakov line
$\Omega_L(\vec{x}) \equiv tr\left(\prod_{t=1}^{L_t}U_0(\vec{x},t)\right)$
 satisfies
\begin{equation}
\Omega_L(\vec{x}+ L_i\vec{e}_i) = \Omega_L^*(\vec{x})\mbox{ for }i=x,y,z
\end{equation}
because of the $C-$periodic boundary conditions.
Therefore, no bulk configurations in either of the two deconfined phases that
have nonvanishing imaginary part of
$\Omega_L\equiv 1/(L^2L_z)\sum_{\vec{x}} \Omega_L(\vec{x})$
can exist and the probability distribution $P_L(\rho)d\rho$ of
$\rho\equiv Re\,\Omega_L$ takes the form sketched in Fig.\ref{fig:pmc}.
The system is most likely in either the confined phase at $\rho^{(1)}$ or
the one remaining deconfined phase corresponding to $\rho^{(2)}$. When
$\rho$ is increased from $\rho^{(1)}$, bubbles of deconfined phase form. These
configurations are suppressed by the interfacial free energy
$\sigma_{cd} A$, where  $A$ is the surface area of the bubble. The largest
bubble  grows until finally $A$ is larger than the area $L^2$ of
two planar  interfaces which devide the lattice into three parts as
depicted in the second part of Fig.~\ref{fig:config}.
Since the interface area of the two planar interfaces is independent of
$\rho$ the probability $P_L$ is constant in the region where their
contributions dominate, i.e. around $\rho^{(min)}$.
Because of the $C-$periodic boundary conditions these interfaces
always separate a region in the confined phase from one in the deconfined
phase that has $Im(\Omega_L)=0$. Thus the corresponding configurations will be
exponentially suppressed by the interfacial free energy of two
confined-deconfined interfaces.

In addition there will be internal fluctuations of
the interfaces. These fluctuations may be described by a simple capillary wave
model (see~\cite{priv92b:histnp}). There one assumes that the interfaces are
almost flat, i.e.  fluctuations  are small. The energy is proportional to the
area of the interface. Expanding the energy around its equilibrium position
leads to a simple Gaussian model for the interface, the so called
``capillary wave model''. In this model the width of an interface increases
only with $\sqrt{\ln\,A}$~(see~\cite{bri86:histnp}). On the other hand, for
configurations corresponding to the center of the histogram the two interfaces
are separated by roughly $L_z/2$.
Therefore, if one choses $L_z\ge L$, the overlap of the two
interfaces for $C$-periodic boundary conditions should be negligible for $L$
large enough. However, it
 turns out that compared to cubic lattices the choice $L_z>L$ will
reduce the interactions, especially for smaller systems. Still,
one should not be in  the region $L_z\gg\xi_t$ where $\xi_t$ is the
tunneling correlation length \cite{gro92b:histesp,bor92:histnp,bor92b:histnp}.
For two independent interfaces one gets \cite{bun92:histesp}
\begin{equation}
P_L^{int}\propto L_z^2\cdot L^{d-3}\cdot
                  \exp\left( -2\sigma_{cd}L^{d-1}\right).
\label{eq:capwav}
\end{equation}
for $d-$dimensional spatial volumes. It was shown in \cite{wie92:histesp} for
$d=2$ that the constraint $\rho=\rho^{(min)}$ does not change this result.

In order to extract $P_L^{int}$ from the total probability distribution
$P_L(\rho)\, d\rho$ we make  an ansatz analogous to the one used in
ref.~\cite{cha86:histnp}: We assume that the total probability distribution
for a system of volume $V=L^2\times L_z$ is given by the sum of two Gaussians
\begin{equation}
P_L^{(i)}(\rho)=
\sqrt{ \frac{V}{2\pi\chi_i}  }\cdot
      \exp\left( -\frac{V}
                      {2\chi_i}(\rho-\rho^{(i)})^2\right),\,
  i=1,2,
\label{eq:plbulk}
\end{equation}
and the interface contribution by eq.~(\ref{eq:capwav}) so that
\begin{equation}
P_L(\rho)\propto
  P_1\,P_L^{(1)}(\rho)+ P_2\,P_L^{(2)}(\rho) + P_3\, P_L^{int}
\label{eq:pltot}
\end{equation}
where $P_1,\, P_2$ and $P_3$ are the relative weights of the two bulk phases
and the interface configurations. In \cite{bor91:histnp} it was shown that
\begin{equation}
P_1  =  \frac{\exp(-\beta Vf_1(\beta,j))}
         { \exp(-\beta Vf_1(\beta,j))+ q\exp(-\beta Vf_2(\beta,j))}
\nonumber
\end{equation}
and
\begin{equation}
P_2  = \frac{q \exp(-\beta Vf_2(\beta,j))}
        {  \exp(-\beta Vf_1(\beta,j)) + q\exp(-\beta Vf_2(\beta,j)) }
\label{eq:p12}
\end{equation}
for a $q-$states Potts model. Here $f_1$ and $f_2$ are the free energy
densities of the two phases. In appendix~\ref{app:fss} we will derive
corresponding finite
size scaling formulas for the Polyakov line and its susceptibility as well as
for some finite volume estimates of the critical coupling $\beta_c$.
We will use these results as a consistency check for this ansatz and  in order
to determine $\beta_c$.

By analogy we conclude that
\begin{equation}
P_3=\frac{\exp(-\beta(V_1f_1+V_2f_2))} {\exp(-\beta Vf_1)+q\exp(-\beta Vf_2)}
\label{eq:p3}
\end{equation}
for the relative weight of a configuration in which  two interfaces separate
regions of volume $V_1$ and $V_2$ in either of the bulk phases from each
other.  Since at the minimum $\rho^{min}$ one has
$V_1\approx V_2\approx V/2$, the dependence on $f_i$ will cancel in the
combination
\begin{equation}
\frac{\overline{P_L}^{max}}{P_L^{min}}\equiv
\frac{\sqrt{P_L^{max,1}\,P_L^{max,2}}}{P_L^{min}}
\label{eq:pmaxmin}
\end{equation}
of the minimal probability $P_L^{min}\propto P_3\,P_L^{int}$ and the maxima
$P_L^{max,i},\,i=1,2$.
This ratio is therefore only weakly dependent on $\beta$. In contrast to this,
in $P_L^{min}\propto P_3\cdot P_L^{int}$ the volume dependence of $P_3$ will
only cancel when $f_1=f_2$, which introduces a fine tuning problem for
$\beta$. In addition, the overall normalization of $P_L(\rho)$ is  needed
for $P_L^{min}$,
 but not for the combination (\ref{eq:pmaxmin}). Thus the latter
is our preferred quantity for the analysis of our numerical data. In order to
cancel the preexponential factors we define
\begin{equation}
 F_L^{(1)}  \equiv  \frac{1}{2 L^{2}}
            \ln\frac{\overline{P_L}^{max}}{P_L^{min}}
            +\frac{3}{4}\frac{\ln L_z}{L^2}
            - \frac{1}{2}\frac{\ln L}{L^2}    .
\end{equation}
This approaches  asymptotically
\begin{equation}
 F_L^{(1)} \rightarrow   \sigma_{cd} + \frac{b_1}{L^2} \vspace{1cm}
              \mbox{ for } L_z\ge L \rightarrow \infty
\label{eq:fl1}
\end{equation}
while
\begin{eqnarray}
 F_L^{(2)} & \equiv & -\frac{1}{2 L^{2}} \ln P_L^{min}
            + \frac{\ln L_z}{L^2}                      \\
           & \rightarrow& \sigma_{cd} + \frac{b_2}{L^2} \vspace{1cm}
              \mbox{ for } L_z\ge L \rightarrow \infty
\label{eq:fl2}
\end{eqnarray}

In order to calculate the probability distribution $P_L(\rho)$, one has to
simulate the SU(3) pure gauge theory at the deconfinement phase transition. But
because of eq.~(\ref{eq:capwav}) any standard local updating algorithm will
have autocorrelation times $\tau_L$ which increase exponentially with $L^2$
("supercritical slowing down"). The use of the multicanonical
algorithm~\cite{ber91a:hist,ber91c:hist,tor77,mar92}
reduces this effect considerably.

\section{The Multicanonical Algorithm}
\label{multica}
Monte Carlo simulations with a local Metropolis or heat bath algorithm
suffer from supercritical slowing down close to first order phase
transitions , i.e. the tunneling time $\tau_L$ for
a system
of volume $V=L^2\times L_z$ is expected to be proportional
to the inverse of the probability of a system with  interfaces
(actually there will be two interfaces because of the boundary conditions),
\begin{equation}
\tau_L\propto V^z\cdot P_L^{min}
      \approx V^z\exp (2 L^{2}\sigma_{cd} ),\vspace{0.5cm}
\end{equation}
with an unknown exponent $z$, and therefore diverges exponentially with the
area $A=L^{2}$ of an interface.

In order to overcome this problem, the multicanonical algorithm does
not sample
the configurations with the canonical Boltzmann weight
\begin{equation}
{\cal P}_L^{can}(S) \propto \exp (\beta S),
\end{equation}
where $S=1/3\,\sum_\Box\,tr\,U_\Box$ is the Wilson action in four
dimensions,
but rather with a modified weight
\begin{equation}
{\cal P}_L^{mc}(S) \propto \exp (\beta_L(S) S+\alpha_L(S)).
\label{eq:boltzmc}
\end{equation}
The coefficients $\alpha_L$ and $\beta_L$ are chosen such that the
probability $P_L$ (not to be confused with the Boltzmann weights) is
increased
for all values of the action in between the two maxima $\rho^{(1)}\mapsto
S_L^{max,1}$ and $\rho^{(2)}\mapsto S_L^{max,2}$ , as shown schematically
in Fig.~\ref{fig:pmc} (where $\rho$ is identified with the action here).
Finally data are analyzed from the multicanonical samples by
reweighting with
$\exp (  (\beta - \beta_L(S)) S  -\alpha_L(S))$.

In order to approximate the weights ${\cal P}_L^{mc}$ leading to the
distribution $P_L^{mc}$ of Fig.~\ref{fig:pmc}, we start from
some good estimate of the canonical distribution $P_L$ (see below) at the
coupling $\beta_P(L)$ which corresponds to equal weight in both phases.
This will lead to different heights of the two peaks.
Then we take a partition
\begin{eqnarray}
 S_L^0 =-6V L_t & < & S_L^1\equiv S_L^{max,1}< S_L^2 < \ldots <
S_L^{N/2}\equiv S_L^{min} < \ldots \nonumber\\& \ldots & < S_L^{N-1}
\equiv S_L^{max,2}
< S_L^N=6VL_t
\end{eqnarray}
of the interval $-6V L_t\le S\le 6V L_t$. The coefficients $\alpha_L$
and $\beta_L$
are chosen to be constants $\alpha_L^k$ and $\beta_L^k$ in the intervals
$[ S_L^k,S_L^{k+1} )$ such that $\ln P_L^{mc}$ interpolates the linear
function
\begin{equation}
 \ln P_L^{max,1} + (\ln P_L^{max,2}- \ln P_L^{max,1}) \cdot
   \frac{S-S_L^{max,1}}{S_L^{max,2}-S_L^{max,1}} .
\end{equation}
continuously between the points $S_L^k$ and $S_L^{k+1}$ for $k=1$ to
$N-2$. The
Boltzmann weight is identical to the canonical one in the first and
last interval.
We arrive at
\begin{equation}
 \beta_L^k-\beta = \left\{
 \begin{array}{ll}
      0 & , k=0 \\
      \delta\beta +  \ln\left(\frac{P_L(S_L^k)} {P_L(S_L^{k+1})} \right)
                           / (S_L^{k+1}-S_L^k)
        & ,k=1,\ldots ,N-2 \\
 0 &  ,k=N-1
 \end{array}
                \right.
\end{equation}
where $\delta\beta\equiv \ln\left( \frac{P_L^{max,2}}{P_L^{max,1}}
\right)
                         /(S_L^{max,2} - S_L^{max,1})$, and the
$\alpha_L^k$ are
given by
\begin{equation}
 \alpha_L^{k+1} \equiv \alpha_L^k + (\beta_L^k - \beta_L^{k+1})
S_L^{k+1},\;
                \alpha_L^0 = \delta\beta\cdot S_L^{max,1}.
\end{equation}
The partition $\{S_L^k\}$ which is in principle arbitrary is defined
by demanding
\begin{eqnarray}
 \frac{P_L (S_L^{k+1})}{P_L(S_L^k)} = \left\{
  \begin{array}{ll}
  1/r_1        &,k=1,\ldots,N/2-1 \\
  r_2          &,k=N/2,\ldots,N-2,
  \end{array}
\right.\nonumber\hfill (2.8)
\end{eqnarray}
such that
$r_1^{N/2-1}=P_L^{max,1}/P_L^{min}$ and
$r_2^{N/2-1}=P_L^{max,2}/P_L^{min}$. This
generalizes the formulas given in
\cite{ber91a:hist,ber91c:hist} to the case
$P_L^{max,1}\neq
P_L^{max,2}$ by introducing a $\delta\beta\neq 0$. In addition the
specific choice for $r_1$ and $r_2$  assures that the
probability of $S_L^{min}$ is lifted by the correct amount.

We apply this algorithm to the $SU(3)$ pure gauge theory  at the
deconfinement phase transition. The multicanonical data sampling was
done with a 5-hit Metropolis algorithm as well as with a Creutz heat
bath algorithm modified according to eq.~(\ref{eq:boltzmc}). In both cases
 three independent $SU(2)$-subgroups are updated following the idea of
ref.~\cite{cab82:hist}. The modifications needed for the Metroplis
algorithms are straightforward. Because of the dependence of $\alpha_L$ and
$\beta_L$ on the total action $S$, the update of the active link has
to be done in scalar mode. However, most of the computation time is
needed for the calculation of the staples surrounding the active link,
and this is still vectorizable.

The modifications for the Creutz algorithm are more complicated. The reason is
that one has to know the action for the updated configuration already for the
proposal of the link. Therefore, whenever it is possible to cross one of the
interval boundaries (e.g.~$S_L^k$) of the multicanonical action, one first has
 to calculate the probabilities $P_{k-1}$ and $P_k$ that the new action
will be below or above the interval boundary.

For simplicity we will describe the modifications for an SU(2) gauge group. The
imbedding in the full SU(3) group is done according to \cite{cab82:hist}.
We adapt the standard notation for the subgroup update (see
e.g.~\cite{ken85:histnp}). Let $h$ be the `active' link, $\Sigma=\xi u,\,u\in
SU(2),$ the sum of the surrounding staples, and
$a=uh= a_0\cdot 1+\vec{a}\cdot\vec{\sigma}$.
Let finally $b$ represent all other
links.  Then one wants to generate a new link $a_0$ with the weight
\begin{equation}
{\cal P}(a_0,b)={\cal N}^{-1}\, \sqrt{1-a_0^2}\,
          \exp\left(\alpha(S)+\beta(S)S(a_0,b)\right) d\,a_0,
                                  \vspace{0.5cm}-1\le a_0\le 1.
\end{equation}
Now $S(a_0,b)$ depends on $a_0$ only through the combination
$s(a_0,\xi)=\frac{2}{3}\xi a_0$. It is convenient to  define
$\tilde{S}(a_0,b)\equiv S(a_0,b)-s(a_0,\xi)$. If $|\tilde{S}-S_L^k|
>\frac{2}{3}\xi$ for
all $k$, no interval boundary can be crossed due to the update of $a_0$,
and one can proceed in the standard way. If on the other hand
$|\tilde{S}-S_L^k|\le\frac{2}{3}\xi$ for one k, the total action $S$
might cross the
interval boundary $S_L^k$ by the update of $a_0$. Then the selection of $a_0$
according to
\begin{equation}
{\cal P}_1(a_0,b)={\cal N}_1^{-1}\,
          \exp(\alpha(S)+\beta(S)S(a_0,b)) d\,a_0,\vspace{0.5cm}-1\le a_0\le 1.
\end{equation}
has to be done in two steps. First one has to select the interval in which the
action will be after the update, and then to select $a_0$ from the
corresponding interval. Let us assume that the intervals
are large enough so that one can only  enter the intervals
$[S_L^{k-1},S_L^k)$ and $[S_L^{k},S_L^{k+1})$. For $SU(3)$ in four dimensions
this means
that the length of each interval must be larger than eight which in practice
is not a very strong restriction. The ratio $r=P_{k}/P_{k-1}$ of the
probabilities to end up in either of the two intervals is given by
\begin{eqnarray}
r&=&\frac{\int_{I_{k-1}}\exp \left(\alpha_{k-1}+ \beta_{k-1}S(a_0,b)\right)}
         {\int_{I_{k}}\exp \left(\alpha_{k}+ \beta_{k}S(a_0,b)\right)}
 \\
 &=& \frac{\beta_{k-1}}{\beta_k}\cdot
     \frac{\exp\left( -\beta_k\left(S_L^{k}- \tilde{S}(a_0,b)-\frac{2}{3}\xi
                \right)\right) -1}
          {1-\exp\left( -\beta_{k-1}\left(S_L^{k}- \tilde{S}(a_0,b)
                    +\frac{2}{3}\xi\right) \right)}  . \nonumber
\end{eqnarray}
Here $I_{k-1}\equiv [-1,3(S_L^k-\tilde{S}(a_0,b))/(2\xi)]$ and
$I_{k}\equiv [3(S_L^k-\tilde{S}(a_0,b))/(2\xi),1]$.
Because of the assumption $P_{k-1}+P_k = 1$ one gets
\begin{equation}
P_{k-1}=\frac{1}{1+r},\vspace{0.5cm}\mbox{ and } P_k=1-P_{k-1}.
\end{equation}
Having chosen the interval with this probability, the selection of $a_0$ is
done again in the standard way by taking
\begin{equation}
a_0=1+\frac{\log x}{\frac{2}{3}\xi}
\end{equation}
where $x$ is  distributed uniformly such that $a_0$ is in the chosen interval.
In a final step one uses an accept/reject procedure to fix up the factor
$\sqrt{1-a_0^2}$.

It is clear that this step cannot be vectorized since the change of any of the
links $b$ might change the couplings $\alpha_L^k$ and $\beta_L^k$ which are to
be taken for the heat bath step for $a_0$. But in a practical application the
total action will hardly ever cross an interval boundary so that one can
allways try to do a number of link updates with the usual, vectorizable Creutz
heat-bath update. One just has to check afterwards whether in any of these
steps one has crossed an interval boundary. The worst which can  happen is
that one has to repeat a few update steps, but as mentioned, this will occur
only rarely. In \cite{gro92:histas} we have demonstrated the
efficiency of the multicanonical algorithm for SU(3) pure gauge theory and
 have shown that one can indeed achieve a vector speed which is comparable
to the canonical Creutz heat bath algorithm.

We apply the algorithm to the determination of the interfacial free energy.

\section{Numerical Results}
\label{results}

\subsection{Fighting the supercritical slowing down}
\label{slowdown}

We simulate systems of size $L^2\times L_z\times 2$ at Wilson couplings
$\beta_{sim}$ close to the deconfinement phase transition. We use a
multicanonical Creutz heat bath algorithm with $N$ interpolating intervals for
the multicanonical action. Each heat bath step was followed by 4 overrelaxation
steps. We define the tunneling autocorrelation time $\tau_L$
by the exponential decay of the autocorrelation function
\begin{eqnarray}
A(t) &=& B(t)/B(0),  \nonumber \\
B(t) &=& \av{ C(k,t) }_{k=1}^{M-t}  \nonumber \\
C(k,t) &=&      (O(k+t) - \av{O(i+t)}_{i=1}^{M-t})
              (O(k) - \av{O(i)}_{i=1}^{M-t})
\end{eqnarray}
with $\av{O(i)}_{i=1}^m := \sum_{i=1}^m O(i)/m $ of the total action $S$. The
autocorrelation time $\tau_L$  is listed in table~\ref{tab:simulations}
together with other simulation parameters. The numbers $\tau_L$ obtained from
$Re(\Omega_L)$ are consistent with this.
\begin{table}[hbt]
\begin{center}
\begin{math}
\begin{array}{|c|c|c|c|c|c|c|c|}    \hline
   L^2\times L_z &  V  & \beta_{sim} & N & N_{upd} & \tau_L &
                 1/P_L^{min} & \overline{P_L^{max}}/P_L^{min}
                  \\   \hline \hline
   8^2\times 8   & 512.00 & 5.0940 & 14 & 56000  &  110(10)
                 & 2.1(3) & 11(1) \\ \hline
   8^2\times 16  & 1024.0 & 5.0947 & 14 & 75000  & 170(10)
                 & 3.3(2) & 19(1) \\ \hline
   8^2\times 24  & 1536.0 & 5.0947 & 14 & 75000  & 190(20)
                 & 1.7(3) & 11(1) \\ \hline
   8^2\times 30  & 1920.0 & 5.0947 & 14 & 70000  & 110(10)
                 & 1.6(4) & 10(1) \\ \hline
   10^2\times 10 & 1000.0 & 5.0950 & 14 & 80000  & 190(30)
                 & 5.9(4) & 37(2) \\ \hline
   10^2\times 30 & 3000.0 & 5.0947 & 14 & 84000  & 500(50)
                 & 9.0(9) & 81(4) \\ \hline
   10^2\times 40 & 4000.0 & 5.0944 & 14 & 88000  & 450(70)
                 & 5.2(5) & 55(5) \\ \hline
   12^2\times 12 & 1728.0 & 5.0928 & 14 & 120000 & 360(60)
                 & 18.3(6)& 100(15) \\ \hline
   12^2\times 36 & 5184.0 & 5.0943 & 14 & 85000 & 1100(120)
                 & 49(3)  & 652(30) \\ \hline
   14^2\times 42 & 8232.0 & 5.0943 & 32 & 115000 & 1300(150)
                 & 290(10)& 4600(200) \\ \hline
\end{array}
\end{math}
\end{center}
  \caption{The autocorrelation time $\tau_L$ in comparison to $P_L^{min}$ and
$\overline{P_L^{max}}/P_L^{min}$.
We did simulations with $N_{upd}$ sweeps of an overrelaxed Creutz heat bath
algorithm at the Wilson coupling $\beta_{sim}$.  $N$ is the number of
interpolating intervals for the multicanonical action.}
  \label{tab:simulations}
\medskip\noindent
\end{table}
Since for the largest lattices we could not observe any tunnelings within
reasonable simulation times using a canonical heat bath algorithm, we cannot
give a quantitative comparison of the algorithms in this case.
Instead, we give the inverse probability $(P_L^{min})^{-1}$ which is expected
to govern the supercritical slowing down of a canonical algorithm. Its growth
is indeed much larger than the growth of $\tau_L$ for the multicanonical
algorithm. Note however that the parameters for the multicanonical update where
not even optimized in all cases which explains the irregularities in $\tau_L$.

\subsection{Finite size scaling}
\label{fss}

\begin{figure}[hbt]
\vspace{5.8cm}
\caption{ {
The Polyakov line $\rho_L$ and its susceptibility $\chi_L$ as a function of the
Wilson coupling.
}}
\protect\label{fig:polyakov}
\vspace{5.8cm}
\caption{{
The coupling $\beta_m$ as a function of the volume together with a linear fit.
}}
\protect\label{fig:betac}
\end{figure}
\begin{table}[htb]
\begin{center}
\begin{math}
\begin{array}{|c|c|c|c|c|}    \hline
   L^2\times L_z & \beta_N &\beta_P & \beta_m & \chi_L^{max}\cdot 100/V
                                            \\   \hline\hline
 8^2\times 8   & -           &  5.0945(5) & 5.0946(40) &   2.7(5)   \\  \hline
 8^2\times 16  & -           &  5.0943(2) & 5.0942(20) &   2.6(5)   \\  \hline
 8^2\times 24  & -           &  5.0939(2) & 5.0938(20) &   2.3(4)   \\  \hline
 8^2\times 30  & 5.0945(10)  &  5.0945(2) & 5.0946(20) &   2.2(7)   \\  \hline
 10^2\times 10 & -           &  5.0949(1) & 5.0948(30) &   2.8(8)   \\  \hline
 10^2\times 30 & 5.0942(10)  &  5.0944(1) & 5.0944(10) &   2.6(9)   \\  \hline
 10^2\times 40 & 5.0939(10)  &  5.0943(1) & 5.0944(10) &   2.5(1.3) \\  \hline
 12^2\times 12 & -           &  5.0940(3) & 5.0942(30) &   2.7(1.5) \\  \hline
 12^2\times 36 & 5.0942(8)   &  5.0942(1) & 5.0942(8)  &   2.7(1.0) \\  \hline
 14^2\times 42 & -           &  5.0943(1) & 5.0942(6)  &   2.6(1.2) \\  \hline
 \infty        & 5.0941(10)  &  5.0943(1) & 5.0943(5)  &   -        \\  \hline
\end{array}
\end{math}
\end{center}
  \caption{The three estimates $\beta_N,\,\beta_m,\,$ and $\,\beta_P$ of the
critical coupling $\beta_c$ together with the maximum $\chi_L^{max}$
of the specific heat. In the last line the infinite volume extrapolations are
given.}
  \label{tab:betac}
\medskip\noindent
\end{table}
{}From the simulations described above we determine the average Polyakov line
$\rho_L$ and its susceptibility $\chi_L(\beta)$. The results for the largest
volumes are shown in Fig.~\ref{fig:polyakov}.
As shown in the appendix~\ref{app:fss}, the ansatz in eq.~(\ref{eq:pltot})
leads to the
following finite size behaviour: The coupling $\beta_N$ at which $\rho_L$
becomes independent of the volume should approach $\beta_c$ exponentially fast
in $L$. The same should be true for the coupling $\beta_P$ for which both peaks
of the probability distribution have the same weight. For our data both
estimates are independent of $L$ for the five largest volumes within the
statistical errors. Therefore we take the average values as  infinite volume
extrapolations. The difference of the
location $\beta_m$ of the maximum of the susceptibility to $\beta_c$ should be
proportional to $1/V^2$ while the leading contribution of the height of the
maximum grows proportional to $V$. As can be seen from Figs.~\ref{fig:polyakov}
and~\ref{fig:betac} and Table~\ref{tab:betac}, our data are consistently
described by this. For the critical coupling we extract an overall average
\begin{equation}
\beta_c=5.0943(1)
\end{equation}
which is consistent with the result of ref.~\cite{ber91:histnp}.
However, the volume of our largest lattice is about five times bigger than
their largest one (which was $12^3$).

\subsection{The interface tension}
\label{sigma}
We  determine the probability distributions for $L_t=2$ and the spatial
volumes $L^2\times L_z$ with $L=8,\,10,\,12,\mbox{ and }14$
and $L_z=3L$. In addition, we vary $L_z$ around this point and compare
to the results for cubic lattices given in~\cite{gro92:histas}.
Fig.~\ref{fig:interface} shows the real part of
$\Omega_L(z)\equiv 1/L^2\,\sum_{x,y}\Omega_L(x,y,z)$ for a typical
configuration close to $\rho^{(min)}$ on  a $14^2\times 42\times 2$
lattice.
\begin{figure}[htb]
\begin{minipage}{0.49\textwidth}
\vspace{5.8cm}
\caption{ {
$Re\,\Omega_L(z)$ for a $14^2\times 42 \times 2 $ lattice. The dotted lines
indicate the bulk expectation values.
}}
\protect\label{fig:interface}
\end{minipage}
\hfill
\begin{minipage}{0.49\textwidth}
\vspace{5.8cm}
\caption{{
Probability distributions for the real part of the Polyakov line
for the largest values of $L_z$.
}}
\protect\label{fig:distas}
\end{minipage}
\end{figure}
As expected from section~\ref{interfacial}, one can
identify two interfaces between the confined phase and the deconfined phase.
The imaginary part of $\Omega_L$ is always zero. In
Fig.~\ref{fig:distas} the resulting probability distributions are shown.
In contrast to the distributions for cubic volumes  ($L=8,\,10,$ and $12$,
see \cite{gro92:histas}) they all have a region of
constant probability in between the two peaks. In Fig.~\ref{fig:dist10} it is
demonstrated for $L=10$ how a plateau forms when one increases $L_z$. The
small increase of $P_L^{min}$ between $L_z=30$ and $40$ is due to the
translational modes of the interfaces and is consistent with
eq.~(\ref{eq:capwav}), as is demonstrated in Fig.~\ref{fig:flz}.
This supports the
scenario developed in section~\ref{interfacial} for asymmetric lattices while
for small cubic lattices it casts some doubt on its applicability. However,
one should note that we are  not (and must not be) in the region
$L_z\gg\xi_t$ where $\xi_t$ is the tunneling correlation length
\cite{gro92b:histesp,bor92:histnp,bor92b:histnp}. For $L=12\mbox{ and }14$
we did not vary $L_z$ but took the existence of a plateau as a criterion for
the validity of our assumptions.
\begin{figure}[hbt]
\begin{minipage}{0.48\textwidth}
\vspace*{-0.3cm}
\vspace{5.8cm}
\caption{{
Probability distributions for $L=10$.
}}
\protect\label{fig:dist10}
\end{minipage}
\hfill
\begin{minipage}{0.49\textwidth}
\vspace{5.8cm}
\caption{{
The $L_z$-dependence of $F_L^{(1)}$ for $L=8\mbox{ and }10$.
}}
\protect\label{fig:flz}
\end{minipage}
\end{figure}
\begin{figure}
\centerline{
	\epsfysize=8cm
	\epsfbox{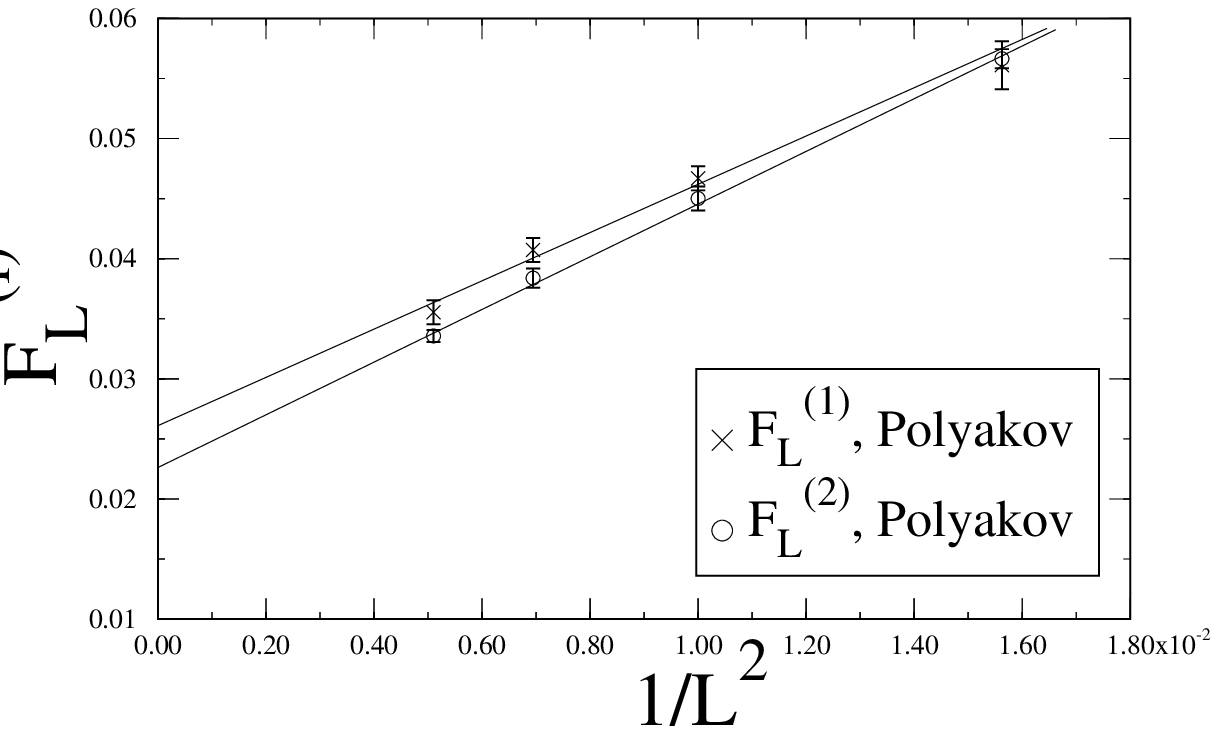}
	}
\caption{{
Results for $F_L^{(i)}$ together with the linear fits.
}}
\protect\label{fig:fl}
\end{figure}

In order to extract the interface tension we evaluate the quantities
$ F_L^{(1)}$ and $ F_L^{(2)}$.
According to eqs.~(\ref{eq:fl1}) and~(\ref{eq:fl2}) both quantities
should be linear functions of
$1/L^2$. Their intercept with the $y$-axis is  $\sigma_{cd}$.
We extract $F_L^{(1)}$ and $F_L^{(2)}$ from the  probability distributions of
Fig.~\ref{fig:distas}. The results are plotted in
 Fig.~\ref{fig:fl} together with the corresponding linear fits. For the
interface tension we get from $F_L^{(1)}$ the value $\sigma_{cd}/T_c^2
= 0.104(7)$. As argued in section~\ref{sigma} the quantity
 $F_L^{(2)}$ is subject to a fine tuning problem.
Nevertheless trying a fit
results in $\sigma_{cd}/T_c^2= 0.090(4)$ which is still rather close to the
other value. Finally we replace the Polyakov line by the total action in all
the calculations and get $\sigma_{cd}/T_c^2= 0.101(6)$ for $F_L^{(1)}$ and
$\sigma_{cd}/T_c^2= 0.092(4)$ for $F_L^{(2)}$. Our overall result taken from
$F_L^{(1)}$ is
\begin{equation}
\frac{\sigma_{cd}}{T_c^2} = 0.103(7)\, \mbox{ for } L_t=2.
\end{equation}
It agrees with the results in
\cite{hua90:hist,kaj91:hist,bro92:hist} (where the error bars are much larger)
 while it is smaller than the one given in~\cite{gro92b:histesp}. This might
indicate that the transverse extensions used in \cite{gro92b:histesp}
(which are at most $8\times8$) are too small, thereby restricting the
fluctuations of the interfaces too strongly. Still the  discrepancy between
these results and \cite{gro92:histas,jan92:hist} which used Binder's histogram
method for cubic volumes is reduced considerably  and can thus be attributed
mainly to interfacial interactions.

Finally we apply this method to the data published for a $16^3\times 4$ and
$28^3\times 4$ lattice in ref.~\cite{fuk89:histnp} and a $12^2\times 24\times
4$ and a $24^2\times 36\times 4$ lattice in ref.~\cite{iwa92:histnp}. Using the
distributions for the absolute value of the Polyakov line we arrive at
\begin{equation}
\frac{\sigma_{cd}}{T_c^2} = 0.040(4)\, \mbox{ for } L_t=4.
\label{eq:sigma4}
\end{equation}
Since only the last two lattices show a reasonable plateau, the systematic
error in this result is certainly still rather large. Simulations on asymmetric
lattices would clarify whether the discrepency to the value $0.027(4)$ given
in~\cite{bro92:hist} is due to this uncertainty. Taking  $\sigma_{cd}$ from
eq.~\ref{eq:sigma4} and the value $L=2.69(3)\, T_c^4$ for $L_t=4$
from ref.~\cite{iwa92:histnp} for the latent heat,
the distance between the nucleation centers would be $R_i=3.8\,$cm according to
eq.~\ref{eq:ri}. This is
probably still too small to survive the proton diffusion. We do not attempt a
fit to the data for $L_t=6$ in ref.~\cite{iwa92:histnp} since for our purposes
there is just one reliable lattice.
\section{Conclusions}
\label{conclusion}

We have applied Binder's histogram method to the determination of the
interface tension in quenched $QCD$. The use of rectangular lattices $L_z>L$
eliminates the interfacial interactions which is reflected by the emergence
of a
plateau in the probability distributions. This modification allows for the
use of smaller $L$-values for the extrapolation to the infinite volume limit.
The number of parameters used for this fit is reduced by the application of the
capillary wave model which describes  our data well. We expect that this will
also improve measurements for other systems like e.~g. the two-dimensional
seven-states Potts model~\cite{jan92:histnp,bor92c:histnp}. In addition, the
use of a multicanonical algorithm  made  simulations possible on much larger
lattices than  before. The value for the interfacial free energy
which we obtain is consistent with the measurements of the Boston and Helsinki
groups although they quote larger errors. Our result  is slightly lower than
the one  obtained by a study of the transfer matrix spectrum. We argue that
this difference might be due to the fact that one is limited to rather small
interfaces in the transfer matrix method.

Since our method neither introduces any pinning forces (like in the older
works) nor requires an exponentially increasing longitudinal extension of the
lattice (like in the transfer matrix method) we believe that it will be very
 efficient  for measurements closer to the continuum limit. At present
the data for $L_t=4$ which were mainly taken on cubic lattices do not show a
clear plateau and should therefore be supplemented by simulations on asymmetric
lattices of the same size.
\section*{Acknowledgement}
We would like to thank U.-J.~Wiese and W.~Janke for valuable discussions. The
simulations were performed on the CRAY-YMP of the HLRZ J\"ulich.

\section*{Appendix}
\appendix
\renewcommand{\theequation}{A.\arabic{equation}}

\section{Finite Size Scaling}
\label{app:fss}
In this appendix we describe a generalization of the results on the finite size
scaling of bulk quantities given in \cite{bor91:histnp} and derive the
corresponding formulas for the interfacial free energy.

We start with the ansatz
\begin{equation}
Z_L(\beta,j) =  \exp(-\beta Vf_1(\beta,j))  + q\exp(-\beta Vf_2(\beta,j))
\label{eq:ztot}
\end{equation}
for the partition function of a system with a temperature driven first order
phase transition close to the transition point. Here $q$ is the number of
ordered phases, $V=L_z\,L^2$ is the volume of the system.
 The ``free energies'' $f_1$ and $f_2$ of the disordered phase 1 and the
ordered phases 2 are assumed to be smooth functions of the coupling
$\beta$ and the external current $j$ which couples to the
order parameter density $\rho$. In \cite{bor91:histnp} it was shown that for
the $q-$state Potts model this relation holds for $q$ large enough and $j=0$.
Corrections are exponentially small in $L$. Relations for the first five
moments of the internal energy
$e_L(\beta)=\frac{1}{V}\frac{d}{d\beta}(\log Z_L(\beta))$
could be obtained by differentiating (\ref{eq:ztot}) with respect to $\beta$.
Assuming that the same is true for nonvanishing $j$, one gets
\begin{eqnarray}
\rho_L(\beta,j)& \equiv & \frac{d}{dj}\ln Z_L(\beta,j) \\
         & = & P_2(\beta,j)\rho_2(\beta,j)\, +\, P_1(\beta,j)\rho_1(\beta,j)\\
         & = & \frac{\rho_1(\beta,j)+\rho_2(\beta,j)}{2} +
               \frac{\rho_1(\beta,j)-\rho_2(\beta,j)}{2}\, th\, Y
\end{eqnarray}
and
\begin{eqnarray}
\chi_L(\beta,j)& \equiv & V\, \frac{d}{dj}\ln \rho_L(\beta,j) \nonumber\\
         & = & \frac{\chi_1(\beta,j)+\chi_2(\beta,j)}{2} +
               \frac{\chi_1(\beta,j)-\chi_2(\beta,j)}{2}\, th\, Y \nonumber\\
         && +  \frac{\left(\rho_1(\beta,j)-\rho_2(\beta,j)\right)^2}{4}\,
               \frac{V}{ch^2 Y}
\label{eq:chil}
\end{eqnarray}
Here we have introduced the probabilities
\begin{equation}
P_1  =  \frac{\exp(-\beta Vf_1(\beta,j))}
           {\exp(-\beta Vf_1(\beta,j)) + q\exp(-\beta Vf_2(\beta,j)) }
\nonumber
\end{equation}
and
\begin{equation}
P_2  = \frac{q \exp(-\beta Vf_2(\beta,j))}
           {\exp(-\beta Vf_1(\beta,j)) + q\exp(-\beta Vf_2(\beta,j))}
\end{equation}
and the expectation values $\rho_i = -\frac{d}{dj}(\beta Vf_i(\beta,j))$ and
susceptibilities $\chi_i = V\,\frac{d}{dj}\left(\rho_i(\beta,j)\right)$.
Furthermore we have used the short hand notation
\begin{equation}
Y = -\frac{\beta V}{2} \left( f_1(\beta,j)-f_2(\beta,j)\right) -
     \frac{\ln q}{2}
\end{equation}
The volume dependence of $\rho_i,\, \chi_i, \mbox{ and }f_i$ is negligible.

Finally, we set $j=0$ and expand all quantities linearly around the critical
$\beta_c$, especially
\begin{equation}
\rho_i(\beta) \approx
             \rho_i - \frac{\tilde{c_i}}{\beta_c^2} (\beta-\beta_c), \, i=1,2
\end{equation}
and
\begin{equation}
Y \approx -V\left( (\beta-\beta_c)\frac{e_1-e_2}{2} -
                   (\beta-\beta_c)^2\frac{c_1-c_2}{4\beta_c^2}\right)
          -\frac{\ln q}{2}
\end{equation}
In order to find the location $\beta_m$ of the maximum of the susceptibility
we take the
derivative of eq.~(\ref{eq:chil}) with respect to $\beta$ and keep only terms
up
to order ${\cal O}((\beta-\beta_c)\cdot V^3)$ and ${\cal O}(V)$. We find
\begin{equation}
\beta_m(L) = \beta_c - \frac{\ln q}{e_1-e_2}\cdot \frac{1}{V}
                     + 8\,(\chi_1-\chi_2)\frac{2(\rho_1-\rho_2)-(e_1-e_2)}
                                              {(\rho_1-\rho_2)^2(e_1-e_2)^2}
                       \cdot\frac{1}{V^2}
\, .
\label{eq:betam}
\end{equation}
where we understand  setting $\beta=\beta_c$ and $j=0$ whenever we omit the
arguments $\beta$ and $j$. Since in our case due to the C-periodic
boundary conditions only one deconfined phase is left, one has to set $q=1$.
 We note a qualitative difference between this case and the case of periodic
boundary conditions (where $q>1$): While in the former one the finite size
corrections are of the form $1/V^2$, they are of the form $1/V$ for the latter
one.

We use these formulas in order to get an estimate for $\beta_c$ and to find a
proper normalization for the probabilities used in the determination of the
interface tension. Following \cite{bor91:histnp} we define the two additional
 finite volume estimates for $\beta_c$: For $\beta_P$ defined via
\begin{equation}
P_L(\rho\le \rho^{min})=P_L(\rho\ge \rho^{min}) \mbox{ for } \beta=\beta_P
\end{equation}
the difference $\beta-\beta_c$ should be exponentially small in $L$.
According to \cite{bor91:histnp}, this should also be true for the point
$\beta_N(V)$ at which
$\rho_L$ becomes independent of $V$.


\begin{thebibliography}{10}
\bibitem{gav89:histnp}
{  R.~V. Gavai, F.~Karsch, and B.~Petersson}, Nucl. Phys. B~{\bf
  322}~(1989)~738.
\bibitem{fuk89:histnp}
{  M.~Fukugita, M.~Okawa, and U.~Ukawa}, Phys. Rev. Lett.~{\bf
  63}~(1989)~1768;  Nucl. Phys. B~{\bf 337}~(1989)~181.
\bibitem{alv90:histnp}
{  N.~A. Alves, B.~A. Berg, and S.~Sanielevici}, Phys. Rev. Lett.~{\bf
  64}~(1990)~3107.
\bibitem{iwa92:histnp}
{  Y.~Iwasaki, K.~Kanaya, T.~Yoshi{\'{e}}, T.~Hoshino, T.~Shirakawa,
  Y.~Oyanagi, S.~Ichii, and T.~Kawai},  Phys. Rev. D~{\bf  46}~(1992)~4657.
\bibitem{kaj86:basf}
{  K.~Kajantie and H.~Kurki-Suonio},  Phys. Rev. D~{\bf  34}~(1986)~1719.
\bibitem{ful88:histnp}
{  G.~M. Fuller, G.~J. Mathews, and C.~Alcock}, Phys. Rev. D~{\bf
  37}~(1988)~1380.
\bibitem{lan93:histnp}
{  K.~Langanke and C.~Rolfs},  Phys. Bl.~{\bf 49}~(1993)~31.
\bibitem{alc89:histnp}
{  C.~Alcock, G.~M. Fuller, G.~J. Matthews, and B.~Meyer},  Nucl. Phys. A~{\bf
  498}~(1989)~301.
\bibitem{ban92}
{  B.~Banerjee and R.~V. Gavai},  Phys. Lett. B~{\bf 293}~(1992)~157.
\bibitem{hua90:hist}
{  S.~Huang, J.~Potvin, C.~Rebbi, and S.~Sanielevici}, Phys. Rev. D~{\bf
  42}~(1990)~2864.
\bibitem{kaj91:hist}
{  K.~Kajantie, L.~K\"arkk\"ainen, and K.~Rummukainen}, Nucl. Phys. B~{\bf
  357}~(1991)~693.
\bibitem{bro92:hist}
{  R.~Brower, S.~Huang, J.~Potvin, and C.~Rebbi},  Phys. Rev.
  D~{\bf 46}~(1992)~2703-2708.
\bibitem{jan92:histnp}
{  W.~Janke},  in Dynamics of First Order Phase Transitions, H.~J. Herrmann,
  W.~Janke, and F.~Karsch, eds., World Scientific (Singapore), 1992, pp.~365,
  reprinted in Int. J. Mod. Phys. C3 (1992) 1137.
\bibitem{jan92b:histnp}
{  W.~Janke}, {\em Multicanonical
  simulation of the two-dimensional 7-state {Potts} model}, in Physics
  Computing, Prague, 1992.
\bibitem{bor92c:histnp}
{  C.~Borgs and W.~Janke},  preprint HLRZ 54-92, FUB-HEP 13/92, J. Physique
  (France) I (in press), 1992.
\bibitem{gro92b:histesp}
{  B.~Grossmann, M.~L. Laursen, T.~Trappenberg, and U.-J. Wiese},
  preprint HLRZ 92-47, 1992, to appear in Nucl. Phys. B;  \\
  {\em The
  confinement interface tension, wetting, and the spectrum of the transfer
  matrix}, to appear in Nucl. Phys. B (Proc. Suppl.)
 (Proceedings of ``Lattice 92'',  Amsterdam), 1992.
\bibitem{bin81:hist}
{  K.~Binder}, Z. Phys. B~43 (1981) 119~; Phys. Rev. A~{\bf  25}~(1982)~1699.
\bibitem{gro92:histas}
{  B.~Grossmann, M.~L. Laursen, T.~Trappenberg, and U.-J. Wiese},
 Phys. Lett. B~{\bf  293}~(1992)~175;
{  B.~Grossmann and M.~L. Laursen},  in Dynamics of First
  Order Phase Transitions, H.~J. Herrmann, W.~Janke, and F.~Karsch, eds., World
  Scientific (Singapore), 1992, pp.~375, reprinted in Int. J. Mod. Phys. C3
  (1992) 1147.
\bibitem{kro91:hist}
{  A.~S. Kronfeld and U.-J. Wiese}, Nucl. Phys. B~{\bf 357}~(1991)~521.
\bibitem{priv92b:histnp}
{  V.~Privman},  in Dynamics of First Order Phase Transitions, H.~J.
  Herrmann, W.~Janke, and F.~Karsch, eds., World Scientific (Singapore), 1992,
  pp.~85, reprinted in Int. J. Mod. Phys. C3 (1992) 857.
\bibitem{bri86:histnp}
{  A.~Bricmont, A.~{El Mellouki}, and J.~Fr{\"o}hlich},  J. Stat.
  Phys.~{\bf 42}~(1986)~743.
\bibitem{bor92:histnp}
{  C.~Borgs and J.~Z. Imbrie}, Comm. Math. Phys.~{\bf 145}~(1992)~235.
\bibitem{bor92b:histnp}
{  C.~Borgs},   Nucl. Phys. B~{\bf 384}~(1992)~605.
\bibitem{bun92:histesp}
{  B.~Bunk},  in Dynamics
  of First Order Phase Transitions, H.~J. Herrmann, W.~Janke, and F.~Karsch,
  eds., World Scientific (Singapore), 1992, pp.~117, reprinted in Int. J. Mod.
  Phys. C3 (1992) 889.
\bibitem{wie92:histesp}
{  U.~J. Wiese},  preprint BUTP-92/37, 1992.
\bibitem{cha86:histnp}
{  M.~S.~S. Challa, D.~P. Landau, and K.~Binder}, Physical Review B~{\bf
  34}~(1986)~1841.
\bibitem{bor91:histnp}
{  C.~Borgs, R.~Koteck\'y, and S.~Miracle-Sol\'e},
 J. Stat. Phys.~{\bf 62}~(1991)~529.
\bibitem{ber91a:hist}
{  B.~A. Berg, U.~Hansmann, and T.~Neuhaus}, preprint FSU-SCRI-91-125, 1991.
\bibitem{ber91c:hist}
{  B.~A. Berg and T.~Neuhaus},  Phys. Lett.B~{\bf 267}~(1991)~249-253;
 Phys. Rev. Lett.~{\bf 68}~(1992)~9.
\bibitem{tor77}
{  G.~M. Torrie and J.~P. Valleau}, J. Comp. Phys.~{\bf 23}~(1977)~187.
\bibitem{mar92}
{  E.~Marinari and G.~Parisi},  Europhys. Lett.~{\bf 19}~(1992)~193.
\bibitem{cab82:hist}
{  N.~Cabibbo and E.~Marinari}, Phys. Lett. B~{\bf 119}~(1982)~387.
\bibitem{ken85:histnp}
{  A.~D. Kennedy and B.~J. Pendleton},  Phys. Lett. B~{\bf
  156}~(1985)~393.
\bibitem{ber91:histnp}
{  N.~A. Alves, B.~A. Berg, and S.~Sanielevici},
   Nucl. Phys. B {\bf 376}~(1992)~218.
\bibitem{jan92:hist}
{  W.~Janke, B.~A. Berg, and M.~Katoot}, Nucl. Phys. B~{\bf382}~(1992)~649.

\end{thebibliography}
%

%
\end{document}